\documentclass[10pt,letterpaper]{article}
\usepackage{opex3}
\usepackage{amsmath}
\usepackage{float}										
\usepackage[latin1]{inputenc}
\usepackage[numbers,sort]{natbib}
\usepackage{hyperref}
\usepackage{booktabs}									
\usepackage{multirow}									
\usepackage{amssymb}
\usepackage{etoolbox}
\apptocmd{\thebibliography}{%
  \setlength{\itemsep}{0pt}%
  \setlength{\parskip}{0pt}%
}{}{}

\newcommand{\ket}[1]{{#1}\rangle}
\newcommand{\bra}[1]{\langle{#1}}
\begin{document}

\title{Utilizing weak pump depletion to stabilize squeezed vacuum states}

\author{T. Denker,$^{1,*}$ D. Sch\"utte,$^{1}$ M. H. Wimmer,$^{1}$ T. A. Wheatley,$^{2}$ E. H. Huntington,$^{3}$ M. Heurs $^{1}$}

\address{$^1$Max-Planck-Institut f\"ur Gravitationsphysik (Albert-Einstein-Institut), \\and Institut f\"ur Gravitationsphysik, Leibniz Universit\"at Hannover,\\Callinstrasse 38, 30167 Hannover, Germany\\$^2$School of Engineering and Information Technology, The University of New South Wales, Australian Defence Force Academy, ACT 2600, Australia\\$^3$The Australian National University, College of Engineering \& Computer Science, ACT 0200, Australia}

\email{$^*$\,timo.denker@aei.mpg.de} 

\homepage{http://quantumcontrol.aei.uni-hannover.de/} 

\begin{abstract}
We propose and demonstrate a pump-phase locking technique that makes use of weak pump depletion (WPD) -- an unavoidable effect that is usually neglected -- in a sub-threshold optical parametric oscillator (OPO). We show that the phase difference between seed and pump beam is imprinted on both light fields by the non-linear interaction in the crystal and can be read out without disturbing the squeezed output. In our experimental setup we observe squeezing levels of $1.96\pm0.01$\,dB, with an anti-squeezing level of $3.78\pm0.02$\,dB (for a 0.55\,mW seed beam at 1064\,nm and 67.8\,mW of pump light at 532\,nm). Our new locking technique allows for the first experimental realization of a pump-phase lock by reading out the pre-existing phase information in the pump field. There is no degradation of the detected squeezed states required to implement this scheme.
\end{abstract}


\hrule
\bibliography{weakpd}

\begin{thebibliography}{10}
\newcommand{\enquote}[1]{``#1''}

\bibitem{Schroedinger1926}
E.~Schrödinger, \enquote{Der stetige {Ü}bergang von der {M}ikro-zur
  {M}akromechanik,} Naturwissenschaften \textbf{14}, 664--666 (1926).

\bibitem{Kennard1927}
E.~Kennard, \enquote{Zur {Q}uantenmechanik einfacher {B}ewegungstypen,}
  Zeitschrift für Physik \textbf{44}, 326--352 (1927).

\bibitem{Darwin1927}
C.~Darwin, \enquote{Free motion in the wave mechanics,} Proceedings of the
  Royal Society of London. Series A \textbf{117}, 258--293 (1927).

\bibitem{Deutsch1985}
D.~Deutsch, \enquote{Quantum theory, the {C}hurch-{T}uring principle and the
  universal quantum computer,} Proceedings of the Royal Society of London. A.
  Mathematical and Physical Sciences \textbf{400}, 97--117 (1985).

\bibitem{Slusher1986}
R.~E. Slusher, L.~W. Hollberg, B.~Yurke, J.~C. Mertz, and J.~F. Valley,
  \enquote{Observation of squeezed states generated by four-wave mixing in an
  optical cavity,} Phys. Rev. Lett. \textbf{56}, 788--788 (1986).

\bibitem{Caves1981}
C.~M. Caves, \enquote{Quantum-mechanical noise in an interferometer,} Phys.
  Rev. D \textbf{23}, 1693--1708 (1981).

\bibitem{Vahlbruch2010}
H.~Vahlbruch, A.~Khalaidovski, N.~Lastzka, C.~Gr{\"a}f, K.~Danzmann, and
  R.~Schnabel, \enquote{The {GEO} 600 squeezed light source,} Classical and
  Quantum Gravity \textbf{27}, 084027 (2010).

\bibitem{vahlbruch2007}
H.~Vahlbruch, S.~Chelkowski, K.~Danzmann, and R.~Schnabel, \enquote{Quantum
  engineering of squeezed states for quantum communication and metrology,} New
  Journal of Physics \textbf{9}, 371 (2007).

\bibitem{Ekert1991}
A.~K. Ekert, \enquote{Quantum cryptography based on {B}ell's theorem,} Phys.
  Rev. Lett. \textbf{67}, 661--663 (1991).

\bibitem{Bennett1993}
C.~H. Bennett, G.~Brassard, C.~Cr\'epeau, R.~Jozsa, A.~Peres, and W.~K.
  Wootters, \enquote{Teleporting an unknown quantum state via dual classical
  and {E}instein-{P}odolsky-{R}osen channels,} Phys. Rev. Lett. \textbf{70},
  1895--1899 (1993).

\bibitem{Douan2001}
L.~M. Duan, M.~D. Lukin, J.~I. Cirac, and P.~Zoller, \enquote{Long-distance
  quantum communication with atomic ensembles and linear optics,} Nature
  \textbf{414}, 413--418 (2001).

\bibitem{Eberle2010}
T.~Eberle, S.~Steinlechner, J.~Bauchrowitz, V.~Händchen, H.~Vahlbruch,
  M.~Mehmet, H.~Müller-Ebhardt, and R.~Schnabel, \enquote{Quantum enhancement
  of the zero-area sagnac interferometer topology for gravitational wave
  detection,} Phys. Rev. Lett. \textbf{104}, 251102 (2010).

\bibitem{Drever1983}
R.~Drever, J.~Hall, F.~Kowalski, J.~Hough, G.~Ford, A.~Munley, and H.~Ward,
  \enquote{Laser phase and frequency stabilization using an optical resonator,}
  Applied Physics B \textbf{31}, 97--105 (1983).

\bibitem{Robins2002}
N.~P. Robins, B.~J.~J. Slagmolen, D.~A. Shaddock, J.~D. Close, and M.~B. Gray,
  \enquote{Interferometric, modulation-free laser stabilization,} Opt. Lett.
  \textbf{27}, 1905--1907 (2002).

\bibitem{Hansch1980}
T.~Hansch and B.~Couillaud, \enquote{Laser frequency stabilization by
  polarization spectroscopy of a reflecting reference cavity,} Optics
  Communications \textbf{35}, 441 -- 444 (1980).

\bibitem{Heurs2009}
M.~Heurs, I.~R. Petersen, M.~R. James, and E.~H. Huntington, \enquote{Homodyne
  locking of a squeezer,} Opt. Lett. \textbf{34}, 2465--2467 (2009).

\bibitem{McKenzie2005}
K.~McKenzie, E.~E. Mikhailov, K.~Goda, P.~K. Lam, N.~Grosse, M.~B. Gray,
  N.~Mavalvala, and D.~E. McClelland, \enquote{Quantum noise locking,} Journal
  of Optics B: Quantum and Semiclassical Optics \textbf{7}, S421 (2005).

\bibitem{Grosse2006}
N.~B. Grosse, W.~P. Bowen, K.~McKenzie, and P.~K. Lam, \enquote{Harmonic
  entanglement with second-order nonlinearity,} Phys. Rev. Lett. \textbf{96},
  063601 (2006).

\bibitem{walls2007}
D.~F. Walls and G.~J. Milburn, \emph{Quantum optics} (Springer, 2007).

\bibitem{bachor2004}
H.-A. Bachor and T.~C. Ralph, \emph{A guide to experiments in quantum optics}
  (wiley-vch, 2004).

\bibitem{svelto1998}
O.~Svelto and D.~C. Hanna, \emph{Principles of lasers} (Springer, 1998).

\bibitem{Hemmerich1990}
A.~Hemmerich, D.~McIntyre, D.~S. Jr., D.~Meschede, and T.~H\"{a}nsch,
  \enquote{Optically stabilized narrow linewidth semiconductor laser for high
  resolution spectroscopy,} Optics Communications \textbf{75}, 118 -- 122
  (1990).

\end{thebibliography}
\vspace{2mm}
\hrule
\section{Introduction}
The existence of squeezed vacuum states was first considered in the 1920's by Schr\"odinger \cite{Schroedinger1926}, Kennard \cite{Kennard1927} and Darwin \cite{Darwin1927}. Discussions about the use of squeezed light led to possible applications in high-precision measurements, quantum computing and quantum communication in the 1980's \cite{Deutsch1985}. Around that time first experiments with vacuum squeezed states were realized \cite{Slusher1986}. It was suggested that the sensitivity of laser interferometric gravitational wave detectors (GWDs) could be improved by injection of squeezed vacuum states \cite{Caves1981}. The first permanent implementation of this method was recently realized in the GWD GEO600 in Hannover \cite{Vahlbruch2010}. As existing GWDs use massive mirrors as macroscopic test masses and are sensitive at Fourier frequencies of Kilohertz down to Hertz, the injected squeezed states need to be stabilized on long timescales \cite{vahlbruch2007}. Besides application in the field of gravitational physics, squeezed states of light are of importance in other major research areas such as continuous variable quantum communication and quantum key distribution \cite{Ekert1991}. Quantum key distribution protocols, entangled states and quantum teleportation have been demonstrated and improved over the past twenty years \cite{Bennett1993, Douan2001, Eberle2010}.\\
In highly complex large-scale experiments such as GWDs every subsystem (e.g. modecleaning cavities, second harmonic generation- or optical parametric oscillator-resonators) has to be stabilized individually to enable measurements. Any locking task requires a suitable error signal, and various techniques have been demonstrated where such an error signal is provided via modulation sidebands (dither locking, Pound-Drever-Hall technique) \cite{Drever1983} or modulation-free via slight misalignment of beams (tilt locking) \cite{Robins2002} or polarization (H\"ansch-Couillaud locking, homodyne locking) \cite{Hansch1980, Heurs2009}. Modulation sidebands provide large error signals, but emerging higher-order modulation frequencies and resulting beatnotes can disturb measurements in that frequency range. The necessary slight misalignments of some modulation-free schemes are direct loss channels for squeezed light so there is a trade-off between additional losses and the capacity to stabilize a system. The phase of a squeezed vacuum state can be locked by using the asymmetry in the quadrature variances due to quantum noise \cite{McKenzie2005}.\\
In this paper we investigate an effect called \textit{weak pump depletion} (WPD) regarding its usefulness for locking the pump field phase to the intracavity field of a non-classical light source. WPD is an omnipresent side effect of the interaction between seed and pump field in the nonlinear medium and automatically contains information about the phase difference between the two beams. The effect of \textit{full pump depletion} can be used for arbitrarily strong entanglement between the two light fields \cite{Grosse2006}.\\ Exploiting this unavoidable interaction we are able to produce and detect a fully stabilized squeezed vacuum state without degrading the squeezed output field, merely by phase-sensitive detection of the transmitted pump field. We term this novel phase-locking scheme \textit{weak pump depletion locking}.
\section{Weak pump depletion}
\label{sec_theory}
In standard textbooks (e.g. \cite{walls2007}) the pump field passing through a (singly resonant) OPO cavity is typically assumed to be constant, even though small pump field variations due to weak pump depletion are unavoidable. In this section we calculate the influence of WPD on both the error signal for pump-phase locking and on the variance of the (anti-)squeezed output.
\subsection{Cavity dynamics}
Let us consider the situation shown schematically in figure~\ref{setup}: A nonlinear crystal is placed inside a folded four mirror bow-tie cavity. The Hamiltonian describing the second-order nonlinear interaction of the cavity modes at fundamental (displayed by the annihilation and creation operators $\hat{a}$ and $\hat{a}^{\dagger}$) and second harmonic frequency ($\hat{b}$ and $\hat{b}^{\dagger}$, respectively) is given by \cite{walls2007}:
\begin{equation}
\textit{$\hat{H}$}=i\hbar\epsilon_\text{c}\left(\hat{b}^\dagger\hat{a}^2-\hat{a}^{\dagger 2}\hat{b}\right),
\label{eqhamiltonian}
\end{equation}
with $\epsilon_\text{c}$ as the nonlinear coupling parameter. Using the cavity equations of motion this leads to the following differential equations (according to \cite{bachor2004}):
\begin {subequations}
\begin{equation}
\dot{\hat{a}} = -2\epsilon_\text{c}\hat{a}^{\dag}\hat{b} - \left(\kappa_\text{a}+i\Delta_\text{a}\right) \hat{a}+ \sqrt{2\kappa_{\text{A}}}\hat{A}_{\text{in}} +\sqrt{2\kappa_{\text{l,A}}}\hat{A_\text{l}},
\end{equation}
\begin{equation}
\dot{\hat{b}} = \epsilon_\text{c}\hat{a}^{2}-\left(\kappa_\text{b}+i\Delta_\text{b}\right)\hat{b} + \sqrt{2\kappa_\text{B}}\hat{B}_{\text{in}} + \sqrt{2\kappa_{\text{l,B}}}\hat{B}_\text{l},
\end{equation}
\end{subequations}
where $\kappa_\text{a}$ and $\kappa_\text{b}$ are the total resonator decay rates for each field (defined in half width half maximum, HWHM) by $2\kappa_\text{a}=-c/l\cdot\mathrm{ln}\left[R_1\cdot R_2\cdot R_3\cdot R_4\cdot(1-L)\right]$ (cf. \cite{svelto1998}) where $L$ is the intracavity loss, $l$ the cavity length, $c$ the speed of light and $R_i$ are the power reflectivities of the mirrors forming the bow-tie cavity (see section \ref{sec_experiment}). $\hat{A}_{\text{in}}$, $\hat{B}_{\text{in}}$, $\hat{A}_{\text{l}}$ and $\hat{B}_{\text{l}}$ are the driving fields with respective coupling rates $\kappa_\text{A}$, $\kappa_\text{B}$, $\kappa_\text{l,A}$ and $\kappa_\text{l,B}$ for input and roundtrip loss experienced by the fundamental and second harmonic light field. All operators are now decomposed into mean value (DC term) and fluctuations (AC term) ($\hat{a} = \alpha + \delta a$, $\hat{A}_{\text{in}} = \alpha_{\text{in}} + \delta A_{\text{in}}$, $\hat{A}_{\text{l}} = \alpha_{\text{l}} + \delta A_{\text{l}}$, $\hat{b} = \beta + \delta b$, $\hat{B}_{\text{in}} = \beta_{\text{in}} + \delta B_{\text{in}}$ and $\hat{B}_{\text{l}} = \beta_{\text{l}} + \delta B_{\text{l}}$). Assuming that the cavity frequency is locked (detuning $\Delta=0$) all detuning terms vanish. The phase angle $\theta_{\text{b}}$ represents the phase difference between the input pump field ($\beta_{\text{in}}$) and the input cavity field ($\alpha_{\text{in}}$) such that $\beta_{\text{in}}=|\beta_{\text{in}}|\text{exp}(i\theta_{\text{b}})$. For simplicity we will drop the `hat formalism' indicating operators for all following calculations.\\ 
As the cavity is singly resonant for 1064\,nm (and not for 532\,nm) we can assume  $\kappa_\text{a} \ll \kappa_\text{b}$, so the pump field interacts with the cavity on a much shorter time scale than the fundamental field (for our considerations $\hat{b}$ is in steady state). Further we only keep terms to first order in $\epsilon_\text{c}$ and $\delta$, and we consider without loss of generality that $\alpha_{\text{in}}$ is real ($\alpha_{\text{in}}^*=\alpha_{\text{in}}$). The steady state intra-cavity field amplitudes can then be written as:
\begin {subequations}
\begin{equation}
\alpha = \frac{\sqrt{2\kappa_\text{A}}\alpha_\text{in}}{\kappa_\text{a}}-\frac{\sqrt{2\kappa_\text{A}}\alpha_\text{in}|\chi|}{\kappa_\text{a}^2}
\label{eqa}
\end{equation}
\begin{equation}
\beta = \frac{\sqrt{2\kappa_\text{B}}|\beta_\text{in}|}{\kappa_\text{B}}e^{i\theta_{\text{b}}}+\frac{\kappa_\text{A}\alpha_\text{in}^2|\chi|}{\sqrt{2\kappa_\text{B}}|\beta_\text{in}|\kappa_\text{a}^2}e^{-i\theta_{\text{b}}}
\label{eqb}
\end{equation}
\end{subequations}
In equations (\ref{eqa}) and (\ref{eqb}) we have introduced the nonlinearityfactor $|\chi|$ by the substitution $\chi=2\epsilon_\text{c}\sqrt{\frac{2}{\kappa_\text{b}}}\beta_{\text{in}}\text{exp}(i\theta_{\text{b}})=|\chi|\text{exp}(i\theta_{\text{b}})$.
$|\chi|$ can be calculated from the value of maximum gain (initial squeezing), which in turn can be calculated from a pair of measured squeezing and anti-squeezing values (see section \ref{chimeasurement}).\\
The outcoupled light fields $\alpha_\text{out}$ and $\beta_\text{out}$ can be calculated using the boundary conditions $\alpha_\text{out}=\sqrt{2\kappa_\text{A}}\alpha-\alpha_\text{in}$ and $\beta_\text{out}=\sqrt{2\kappa_\text{B}}\beta-\beta_\text{in}$ (and writing the field quadratures as $X_\text{q}^+=(q+q^*)$ and $X_\text{q}^-=i(q-q^*)$ for $q=\alpha_{\text{out}}, \beta_{\text{out}}$) \cite{bachor2004}:
\begin {subequations}
\begin{equation}
X_{\alpha_{\text{out}}}^+=\frac{2\alpha_\text{in}\left(2\kappa_\text{a}\kappa_\text{A}-\left(\kappa_\text{a}^2-|\chi|^2\right)\right)}{\kappa_\text{a}^2-|\chi|^2}-\boxed{\frac{4\alpha_\text{in}\kappa_\text{A}|\chi|}{\kappa_\text{a}^2-|\chi|^2}\cos{(\theta_{\text{b}})}}
\label{alphaoutp}
\end{equation}
\begin{equation}
X_{\alpha_{\text{out}}}^-=\frac{4\kappa_\text{A}\alpha_{\text{in}}|\chi|}{\kappa_\text{a}^2-|\chi|^2}\sin{(\theta_{\text{b}})}
\label{alphaoutm}
\end{equation}
\end{subequations}
Equations (\ref{alphaoutp}) and (\ref{alphaoutm}) show that $X_{\alpha_{\text{out}}}$ is a function of both $\alpha_{\text{in}}$ and $|\chi|$, which in turn is a function of $|\beta_{\text{in}}|$. The influence of the pump field and the relative phase $\theta_{\text{b}}$ between pump and seed beam appear in the second term of the expression for $X_{\alpha_{\text{out}}}^+$, indicated by the box. This signal can be used as an error signal for locking purposes (see section \ref{sec:error}). In contrast, the effect of the nonlinear interaction between pump and seed on the output pump field is typically ignored (in the so-called ``no pump depletion'' limit). The effect is indeed very small; however, it does not vanish and can be seen in the output field quadratures of the pump field: 
\begin {subequations}
\begin{equation}
X_{\beta_{\text{out}}}^+=\left(2|\beta_{\text{in}}|+\frac{2\alpha_\text{in}^2\kappa_\text{A}|\chi|}{|\beta_{\text{in}}|\kappa_\text{a}^2}+\frac{4\alpha_{\text{in}}^2|\chi|^2\kappa_\text{A}}{|\beta_{\text{in}}|\kappa_\text{a}^3}\right)\cos{(\theta_{\text{b}})}
\label{betaoutp}
\end{equation}
\begin{equation}
X_{\beta_{\text{out}}}^-=\boxed{\left(-2|\beta_{\text{in}}|-\frac{4\alpha_{\text{in}}^2|\chi|^2\kappa_\text{A}}{|\beta_{\text{in}}|\kappa_\text{a}^3}\right)\sin{(\theta_{\text{b}})}}
\label{betaoutm}
\end{equation}
\end{subequations}
Equations (\ref{betaoutp}) and (\ref{betaoutm}) are also functions of $\alpha_{\text{in}}$, $\beta_{\text{in}}$, and their phase relationship. Therefore there will be small fluctuations around the average pump field amplitude caused by the nonlinear interaction with the seed field. This is termed weak pump depletion. Usually neglected, this term shows a sinusoidal dependence on the pump-seed phase relationship $\theta_{\text{b}}$. Under certain constraints (to be examined below) the boxed term of eq. (\ref{betaoutm}) can be used as an error signal to lock the pump-seed phase angle. This angle determines the quadrature of squeezing; locking it will additionally lock the quadrature angle of squeezing illustrated in the quadrature variance in section \ref{sec:sqz}.\\
\noindent Polarization-based homodyne detection \cite{Heurs2009} of the signal in eq. (\ref{betaoutm}) will result in a large DC field that varies as a function of the pump-seed phase angle $\theta_{\text{b}}$. In the case of a singly resonant cavity the s-polarized pump field (\ref{betaoutm}) containing the WPD-signal is mixed by a quarter-wave-plate and a polarizing beam splitter with the local oscillator (LO) (the amount of the pump field in p-polarization $|\beta_{\text{p,in}}|\text{exp}(i(\theta_\text{b}+\gamma))$ that it does not interact with the crystal). The difference in the two detected intensities is  
\begin{equation}
I_{\text{lock}} = \frac{2|\chi|\kappa_A\alpha_{\text{in}}^2}{\kappa_\text{a}^2|\beta_{\text{in}}|}|\beta_{\text{p,in}}|\sin{(\theta_{\text{b}}+\gamma)},
\label{HDlock}
\end{equation}
where $\gamma$ is the homodyne measurement angle. It has a fixed value of $\gamma=\pi/2$ due to the fact that signal and LO have orthogonal polarizations.
\subsection{The effect of WPD on squeezing and anti-squeezing}
\label{sec:sqz}
The results for the fluctuating terms $\delta$ are found by moving to the Fourier frequency domain $\text{FT}\left[\frac{d\delta a(t)}{dt}\right] = - i\omega \text{FT}\left[\delta a(t)\right]$. The variance of $\delta X_{A_{\text{out}}}^\pm$ gives the values for (anti-)squeezing in the amplitude or phase quadrature, respectively. We use the common definition for quadrature variance: $V^{\pm} \equiv \bra{|\delta X^{\pm}|}\ket{}^2 - \bra{|\delta X^{\pm}|^2}\ket{}$, where $\delta X^+_A = \delta A + \delta A^\dag $ and $\delta X^- = i(\delta A - \delta A^\dag) $. As we are only interested in the variance of the (squeezed) output at the fundamental frequency we look for the phase and amplitude quadrature expressions for the seed field ($\delta A_{in}$) at different pump-seed phase angles $\theta_{\text{b}}$:
\begin{flalign}
\begin{split}
V^+_{A_{\text{out}}}(\theta_{\text{b}}=0)&=V^+_{A_{\text{in}}}\frac{\left(|\chi|+\kappa_\text{a}-2\kappa_\text{A}\right)^2}{\left(\kappa_\text{a}+|\chi|\right)^2}+\boxed{V^+_{B_{\text{in}}}\frac{4\alpha_\text{in}^2\kappa_\text{A}^2|\chi|^2}{|\beta_\text{in}|^2\kappa_\text{a}^2\left(\kappa_\text{a}+|\chi|\right)^2}}+V^+_{A_\text{l}}\frac{4\kappa_\text{l,A}\kappa_\text{A}}{\left(\kappa_\text{a}+|\chi|\right)^2}\\&=V^-_{A_{\text{out}}}(\theta_{\text{b}}=\pi/2)
\end{split}
\label{sqzeq1}
\end{flalign}
\begin{flalign}
\begin{split}
V^-_{A_{\text{out}}}(\theta_{\text{b}}=0)&=V^-_{A_{\text{in}}}\frac{\left(|\chi|-\kappa_\text{a}+2\kappa_\text{A}\right)^2}{\left(\kappa_\text{a}-|\chi|\right)^2}+\boxed{V^-_{B_{\text{in}}}\frac{4\alpha_\text{in}^2\kappa_\text{A}^2|\chi|^2}{|\beta_\text{in}|^2\kappa_\text{a}^2\left(\kappa_\text{a}-|\chi|\right)^2}}+V^-_{A_\text{l}}\frac{4\kappa_\text{l,A}\kappa_\text{A}}{\left(\kappa_\text{a}-|\chi|\right)^2}\\&=V^+_{A_{\text{out}}}(\theta_{\text{b}}=\pi/2)
\end{split}
\label{sqzeq2}
\end{flalign}
The boxed terms shows the miniscule influence of the pump field phase on the fundamental field ($\alpha_\text{in}\ll|\beta_\text{in}|$). Our calculations show that there should be no degradation of the detected quadrature variances if we use the highlighted expression in eq. (\ref{betaoutm}) as an error signal for pump phase locking.\\
By neglecting intracavity losses ($\kappa_\text{l,A}=0$ and $\kappa_A = \kappa_a$) and pump field influence the equations (\ref{sqzeq1}) and (\ref{sqzeq2}) turn into the variance of the initial squeezing:
\begin{equation}
V^\pm_{\text{init}}=V^\pm_{A_{\text{in}}}\frac{\left(|\chi|\mp\kappa_\text{a}\right)^2}{\left(\kappa_\text{a}\pm|\chi|\right)^2}
\label{sqzeq3}
\end{equation}
\subsection{Gain and losses}
\label{chimeasurement}
\begin{figure}[H]
				\centering
                \includegraphics[width=10cm]{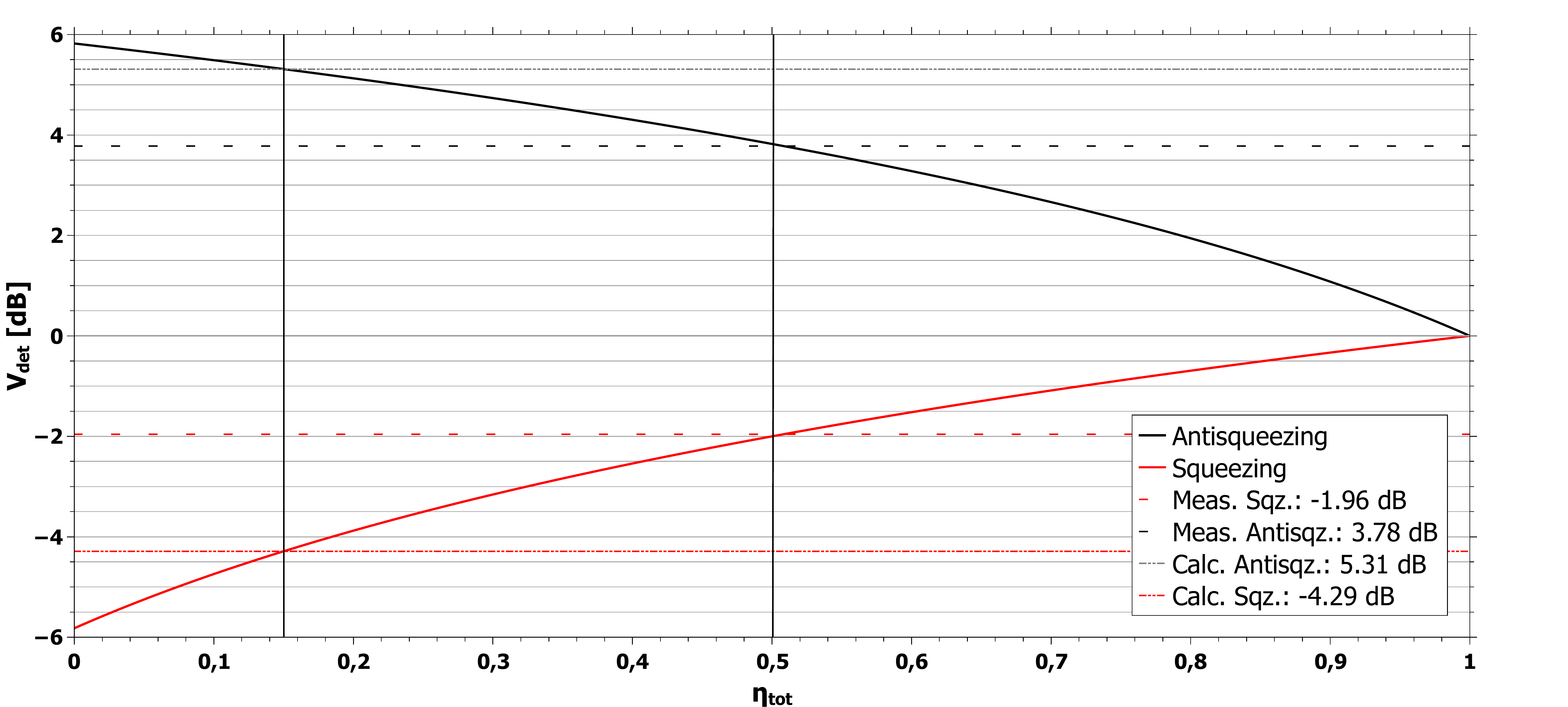}
                \caption{Plot of detected variance (eq. (\ref{losschannel})) as a function of total loss $\eta_\text{tot}$ with the initial squeezing value $V_\text{init}^\pm=\pm 5.82$ dB.}
				\label{sqzloss}
  \end{figure}

We are interested in the nonlinearityfactor $\chi$ related to the maximum amount of \mbox{(anti-)squeezing}. To calculate the initial (anti-)squeezing values $V_\text{init}^\pm$ we assume that the measured values $V_{det}^\pm$ experience identical optical loss $\eta_\text{tot}$. The incoupling loss can be treated like an open beam splitter port. In this case the variance including losses becomes:
\begin{equation}
V_{\text{det}}^{\pm}=\eta_\text{tot} V_{\text{init}}^\pm+\left(1-\eta_\text{tot}\right)
\label{losschannel}
\end{equation}
For this reason we equalize eq. (\ref{losschannel}) for $V_{\text{det}}^+$ and $V_{\text{det}}^-$ with $V_\text{init}^+=-V_\text{init}^-$ for the same loss value $\eta_\text{tot}$ and obtain
\begin{equation}
V_\text{init}^-=-V_\text{init}^+=\frac{V_\text{det}^--1}{V_\text{det}^+-1}
\label{initsqz}
\end{equation}
With the result the total optical loss factor $\eta_\text{tot}=0.5$ and the value for initial squeezing $V_\text{init}^\pm=\pm 5.82$ dB are calculated. Figure \ref{sqzloss} shows a plot of eq. (\ref{losschannel}) for $V_\text{init}^\pm=\pm 5.82$ dB over the loss factor $\eta_\text{tot}$
\noindent
The theoretical characterization of the OPO behavior in equations (\ref{sqzeq1}) and (\ref{sqzeq2}) includes intracavity losses via the decay rate of the intracavity field $\kappa_a$ or $\kappa_A$ for the input field, given by the escape efficiency $\eta_\text{esc}$. Different loss factors degrade the initial squeezing to arrive at the outcoupled squeezing and finally the measured squeezing at the homodyne detector:
$$V^\pm_{\text{init}}\xrightarrow{\eta_{\text{esc}}}V^\pm_{A_\text{out}}\xrightarrow{\eta_\text{prop}\eta_\text{h}\eta_\text{qe}}V^\pm_{\text{det}}.$$ This chain of reasoning shows the connection between all loss terms and the different quadrature variances. With the knowledge of each loss term of $\eta_\text{tot}=\eta_\text{esc}\eta_\text{prop}\eta_\text{h}\eta_\text{qe}$ and the relation of $V^\pm_{\text{det}}$ and $V^\pm_{A_\text{out}}$ (see equations (\ref{losschannel}) and (\ref{initsqz})) the theoretical output $V^\pm_{A_\text{out}}$ can be compared with the experimental data $V^\pm_{\text{det}}$ in \ref{comp} (see figure \ref{sqzslope}).
\section{Experiment}
\label{sec_experiment} 
\begin{figure}[H]
	\centering
	\includegraphics[width=12cm]{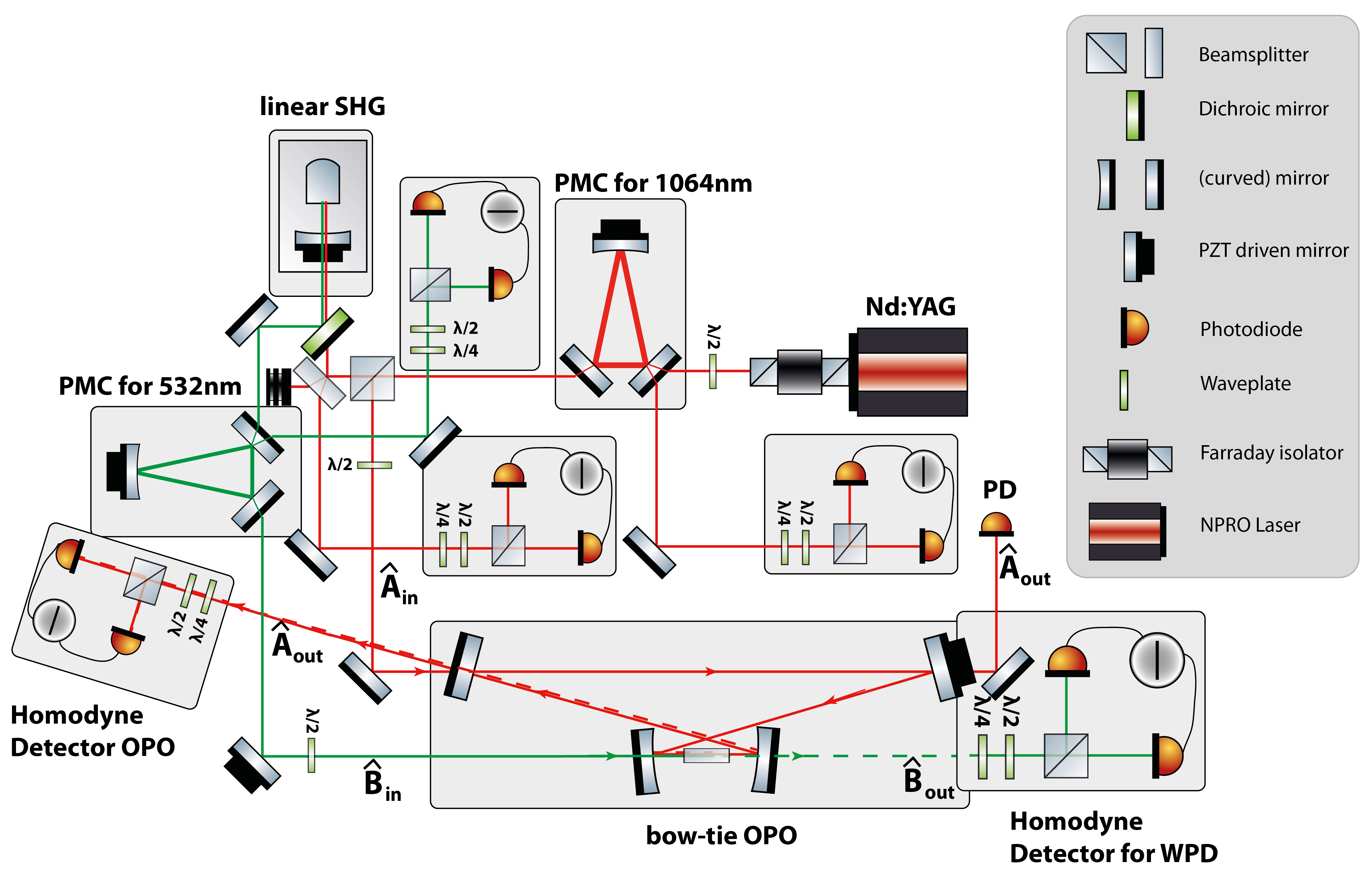}
	\caption{Experimental setup}
	\label{setup}
	\end{figure}

Figure \ref{setup} shows a simplified schematic of our experimental setup. Modematching optics are omitted for clarity. The laser source is a $\lambda=1064$\,nm Nd:YAG non-planar ring oscillator (NPRO) (Innolight Mephisto). To generate the pump light at $\lambda=532$\,nm we use a linear hemilithic SHG with a 6.5\,mm lithium niobate(LiNbO$_3$)-crystal frequency stabilized by polarization based homodyne locking \cite{Heurs2009}. The squeezed vacuum state is generated by a subthreshold bow-tie OPO singly resonant for the fundamental field. It consists of two curved mirrors (radius of curvature 50\,mm) and two flat mirrors. All but one have a high reflective coating ($R=0.999$) for fundamental (1064\,nm) and anti-reflective coating ($R<0.15$) for second harmonic frequency (532\,nm). The first flat mirror acts as in/out-coupler to the cavity (its power reflectivity for 1064\,nm is $R=0.9$\, and it is anti-reflective coated for 532\,nm ($R<0.15$)). One mirror is mounted on a piezo-electric transducer (PZT)-crystal to adjust the cavity length to the laser frequency. A 10\,mm long periodically-poled potassium titanyl phosphate (PPKTP)-crystal is placed between the two curved mirrors to generate squeezing. The temperature quasi-phasematching between seed and pump light is provided by a custom oven design. The beam waist inside the crystal is $\omega_0\approx20\,\mu$m $(1/e^2\,)$. The OPO cavity is also frequency locked with the polarization based homodyne locking scheme \cite{Heurs2009} which allows detection of (anti-)squeezing in the phase quadrature at the same time (via the fixed homodyne measurement angle $\gamma = \pi/2$, see eq. (\ref{HDlock})). The characteristic parameters of the cavity are listed in table \ref{parameters}.
	\noindent
The OPO locking beam has a power of 9.66\,mW with a ratio of 3:100 between signal (s-polarization) and local oscillator (p-polarization). The power of the green pump light is 67.8\,mW with s/p-polarization ratio of 100:5. This slight deviation from the experimental ``rule of thumb'' for homodyne detection of 1:100 is needed to guarantee adequate experimental performance. It provides a sufficiently stable lock while the local oscillator power required for the quadrature homodyne measurement is still high enough. Due to the highly sensitive measurement of the WPD-locking-signal (eq. (\ref{betaoutm})) there is a trade-off between the amount of pump power and attenuation with several neutral density filters to keep the photodiodes operational. By detecting the transmitted green pump light with a homodyne detector we are able to lock its phase to the cavity. The PZT mirror in the pump beam can be used as a phase actuator and also provides the option of inducing a kHz-modulation for dither-locking \cite{Hemmerich1990} of the pump phase for comparison between the two locking schemes (WPD and dither).
\subsection{Error Signals}
\label{sec:error}
\begin{figure}[H]
        \centering
                \includegraphics[width=0.6\textwidth]{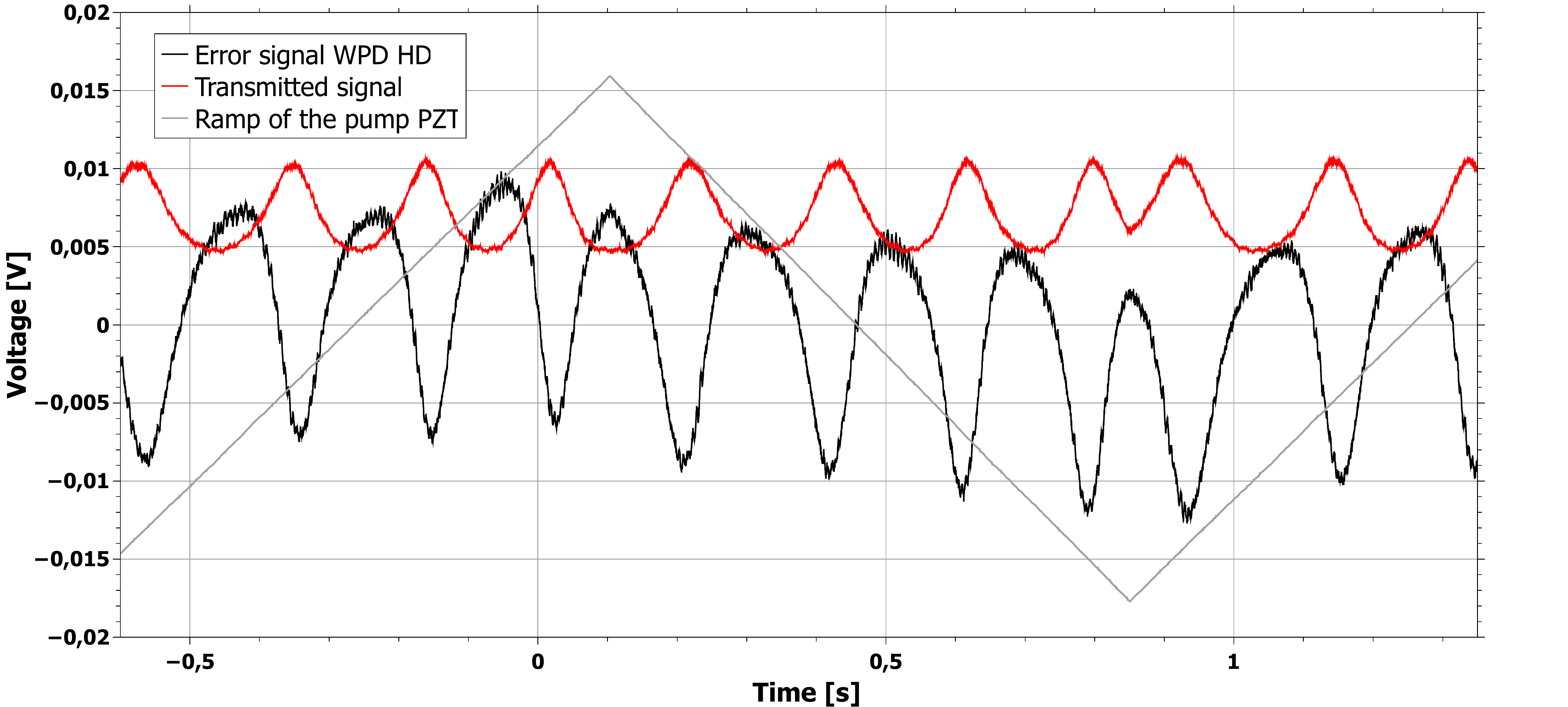}
								\vspace{2mm}
                \caption{Error signal of the transmitted pump light for WPD (black) and transmitted cavity field (red) with swept pump phase versus time. The slight phase offset gives a strong enough signal for pump phase stabilization. For purposes of presentation the scaling of the two signals differs. The tiny effect of WPD on the pump phase is lowpass filtered and amplified by the homodyne photodetector electronic.}
								\label{error}
  \end{figure}

Figure \ref{error} shows the error signal for WPD-lock from eq. (\ref{HDlock}), and in comparison $X_{\alpha_{out}}^+$ from eq. (\ref{alphaoutp}) (the transmitted infrared signal of the OPO which is indicative of the behavior of the green pump light). The amplitude quadrature of the infrared light ($X_{\alpha_{out}}^+$) is detected with a single photodetector (red curve in figure \ref{error}) whereas the homodyne detector acts as an interferometer to read out the phase quadrature of the green pump light ($X_{\beta_{\text{out}}}^-$) (black curve in figure \ref{error}). $X_{\beta_{\text{out}}}^-$ can be used as a locking signal to stabilize the pump phase $\theta_{\text{b}}$, as its gradient is maximal close to the extrema of the infrared light field exiting the cavity $X_{\alpha_{out}}^+$ (deviations are due to delays in the phase detection).
\section{Results}
\label{sec_results}
\begin{figure}[H]
                \includegraphics[width=0.5\textwidth]{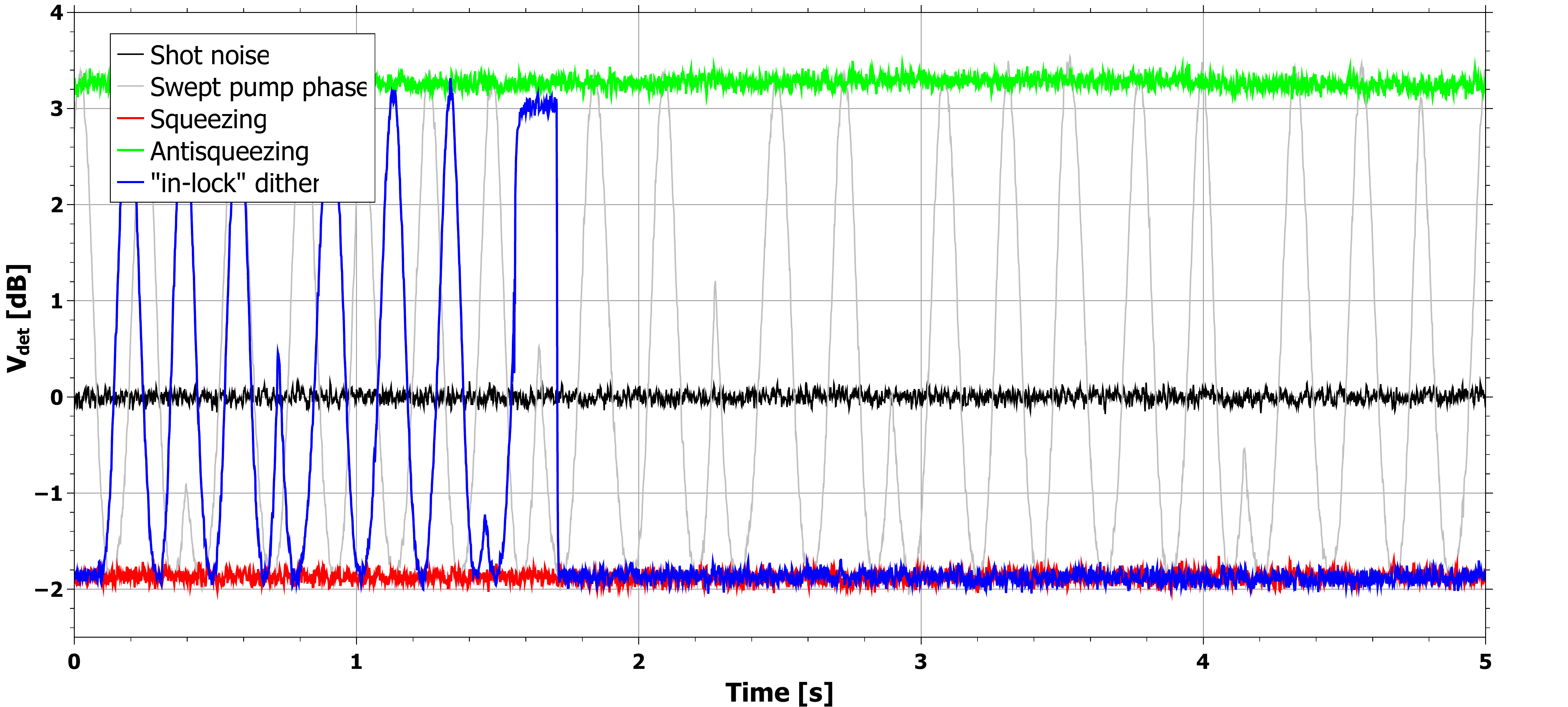}
                \includegraphics[width=0.5\textwidth]{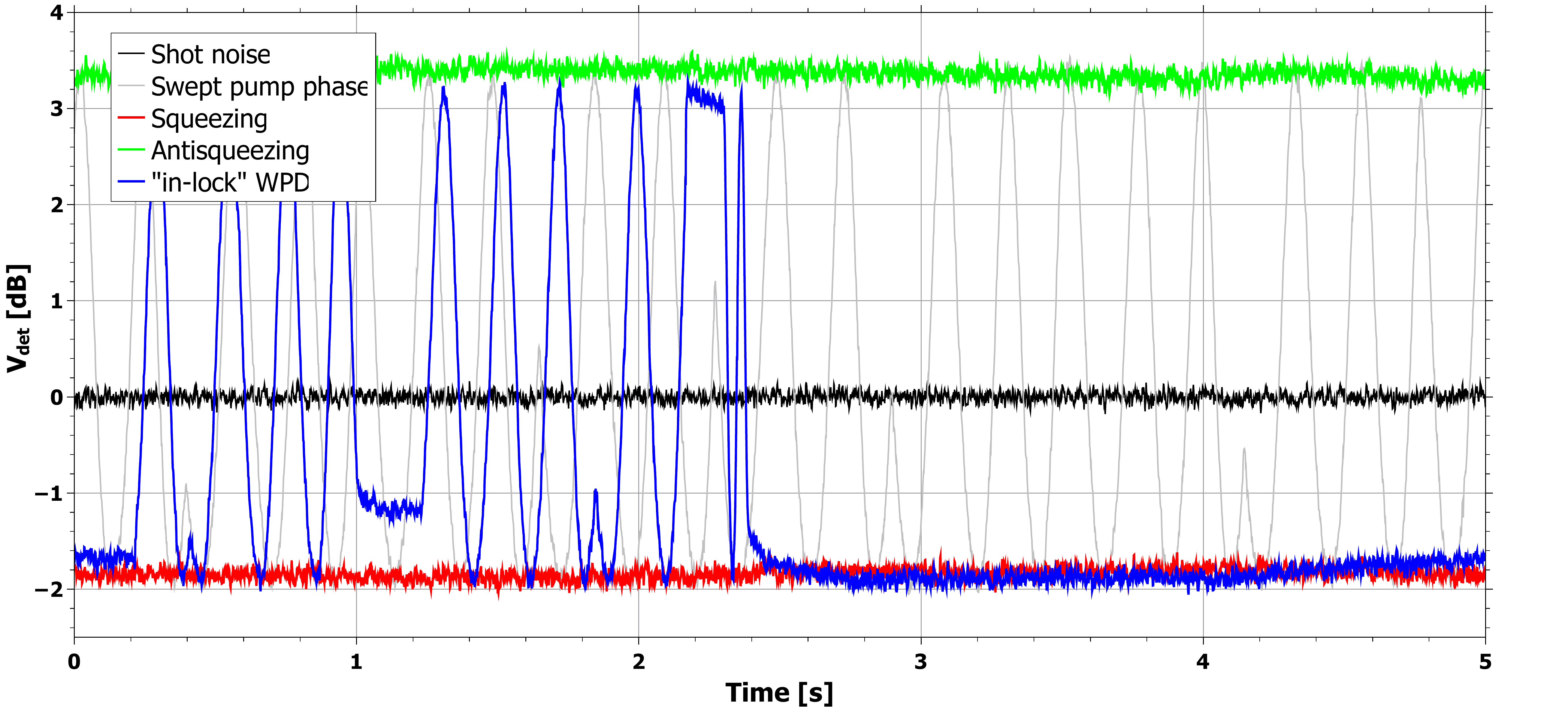}
        \caption{Zero span measurements of the shot noise levels for the two different locking schemes (Left: OPO stabilization via dither locking. Right: OPO stabilization via WPD locking.). The shot noise level without pump is shown in black (with scanned pump phase grey), pump phase locked to anti-squeezed phase quadrature in green, pump phase locked to squeezed phase quadrature in red. The blue line illustrates the behaviour of switching the scanned pump phase to in-lock.}
	\label{zerospan}
  \end{figure}

\begin{figure}[H]
                \includegraphics[width=0.5\textwidth]{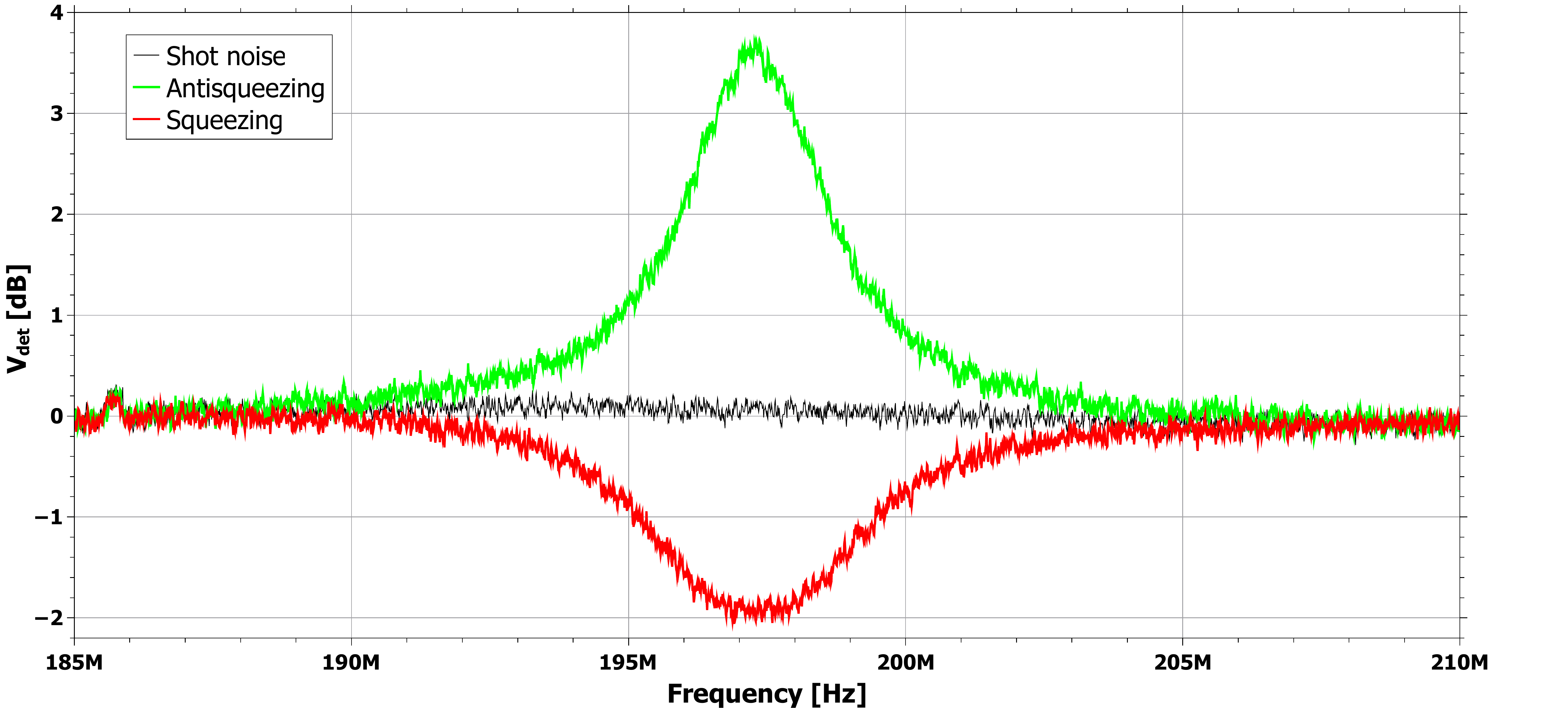}
                \includegraphics[width=0.5\textwidth]{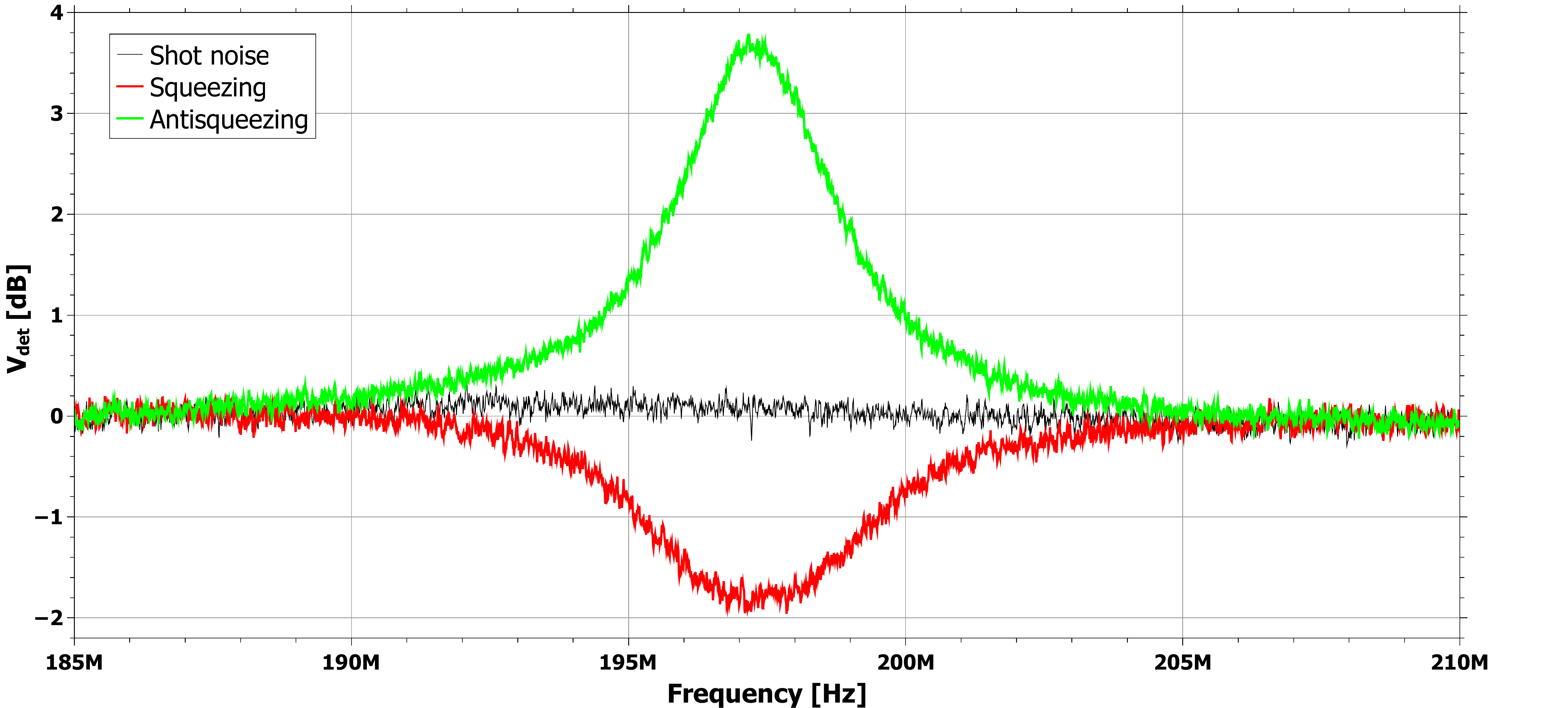}
        \caption{Squeezing spectrum around 197.4\,MHz with squeezing level (red) and anti-squeezing level (green) versus scanned frequency. Left: OPO stabilization via dither locking. Right: OPO stabilization via WPD locking. The spectrum analyzer settings were: 1.5\,MHz resolution bandwidth and 91\,Hz video bandwidth with a sweep time of 1.8\,s and an internal attenuation of 6\,dB. Averaging factor is 10. The noise levels were normalized to shot noise (black).}\label{span}
  \end{figure}

We used a multichannel oscilloscope (Agilent MSO X 2014A) to display the error signal detected by the homodyne photodetector and for the transmitted light detected by a single photodetector. The variance of the squeezed output at the fundamental frequency is recorded on a signal analyzer (Agilent MXA N9020A) when the system was locked.
Figure \ref{zerospan} shows the zerospan measurements at 197.4\,MHz (first free spectral range, FSR) of the variances of the (anti-)squeezed output (see eq. (\ref{sqzeq1}) and (\ref{sqzeq2})) as a comparison between (a) a conventional dither-locking scheme and (b) homodyne locking using weak pump depletion. The swept variance due to the swept pump phase was stabilized at its minimum by switching on the lock. Figure \ref{span} shows the (anti-)squeezing spectra when the pump phase was locked. The observed squeezing levels are $(1.96\pm0.01)$\,dB, with an anti-squeezing level of $(3.78\pm0.02)$\,dB.
	\noindent
The comparison of the measurements shows that WPD- and dither-lock result in the same squeezing levels for identical configuration as predicted by equations (\ref{sqzeq1}) and (\ref{sqzeq2}) for our experimental parameters.
\subsection{Comparison between theory and experiment}
\label{comp}
\begin{figure}[H]
        \centering
                \includegraphics[width=12cm]{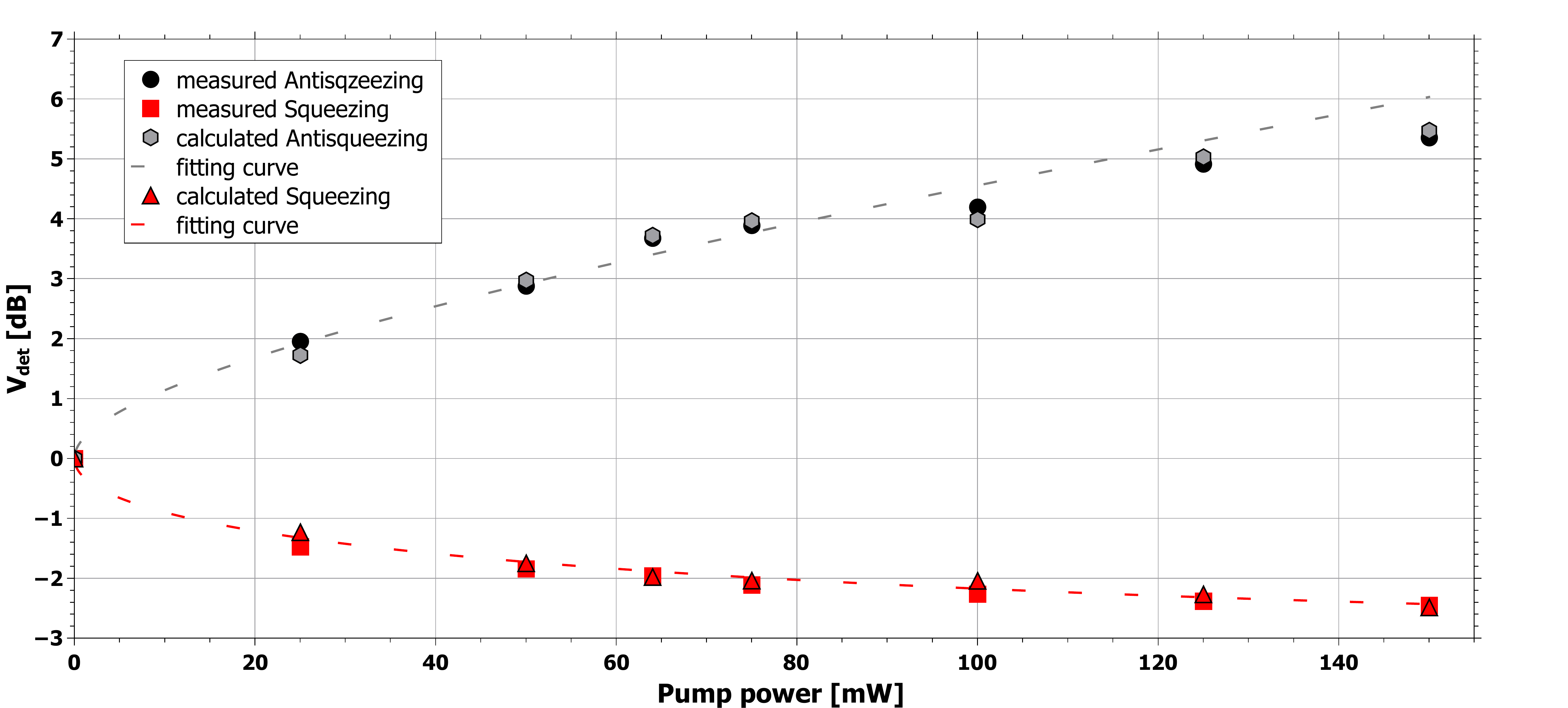}
                \caption{Plot of pump power dependences of the (anti-)squeezing levels by calculating the outcoupled (anti-)squeezing $V^\pm_{A_{\text{out}}}$ (see eq. (\ref{sqzeq1}) and (\ref{sqzeq2})) and taking the loss factor of $\eta_\text{prop}\eta_\text{h}\eta_\text{qe}=0.59$ into account (eq. (\ref{losschannel})).}
	\label{sqzslope}
	\end{figure}

To show the validity of the developed theoretical model we compare the calculated parameters with the measurements. All necessary parameters for the setup are given in table \ref{parameters}. 
\begin{table}[H]
\small
\centering
\caption{Overview of all parameters characterizing the OPO. The calculation of the parameter $\chi$ is described in section \ref{chimeasurement}.}
\begin{tabular}{llll} 
\toprule
parameter&symbol&value&unit\tabularnewline
\midrule
length&$l$&1.52 & m\tabularnewline
free spectral range&FSR&197.4&MHz\tabularnewline
Finesse &$\mathfrak{F}$&58&\tabularnewline
linewidth (FWHM)& $\Delta \nu$&3.9575&MHz\tabularnewline\midrule
\multirow{3}{3.2cm}{cavity decay rates\\(in HWHM)} & $\kappa_\text{a}$ &$1.2434\cdot10^7$& rad/s\tabularnewline
& $\kappa_\text{A}$&$1.0686\cdot10^7$&rad/s\tabularnewline
&$\kappa_{\text{l,A}}$&$0.1749\cdot10^7$&rad/s\tabularnewline\midrule
\multirow{2}{*}{input power (seed)}&\multirow{2}{*}{$\alpha_{\text{in}}^2$}&0.00055&W\tabularnewline
&&$2.946\cdot 10^{15}$&Hz\tabularnewline
\multirow{2}{*}{input power (pump)}&\multirow{2}{*}{$\beta_{\text{in}}^2$}&0.064&W\tabularnewline
&&$2.143\cdot 10^{17}$&Hz\tabularnewline\midrule
nonlinearityfactor (in HWHM)&$|\chi|=2\epsilon_\text{c}\sqrt{\frac{2}{\kappa_\text{b}}}|\beta_{\text{in}}|\text{exp}(i\theta_{\text{b}})$&$3.45\cdot10^6$&rad/s \tabularnewline
initial (anti-)squeezing&$V^\pm_\text{init}$&$\pm$5.82&dB\tabularnewline\midrule
calculated squeezing&$V^+_{A_{\text{out}}}$&-4.29&dB\tabularnewline
calculated anti-squeezing&$V^-_{A_{\text{out}}}$&5.31&dB\tabularnewline\midrule
detected squeezing&$V^+_{\text{det}}$&-1.96&dB\tabularnewline
detected anti-squeezing&$V^-_{\text{det}}$&3.78&dB\tabularnewline
\bottomrule
\end{tabular}
\label{parameters}
\end{table}
\normalsize
\noindent
For our setup we determined that the propagation efficiency is $\eta_\text{prop}=0.92$, the homodyne visibility\footnote{Mismatch efficiency due to mismatches of the TEM00-modes in orthogonal polarizations} $\eta_\text{h}=(1-0.11/1+0.11)^2=0.64$, and the quantum efficiency of the photodiodes $\eta_\text{qe}=0.98$. $\eta_\text{esc}=T/(T+L)=0.85$ is the escape efficiency with intracavity losses\footnote{Recalculated with \cite{svelto1998} of $\kappa_a$ inferred from the measurement of the linewidth of the (lossy) cavity} $L=0.018$ and $T=0.1$ as the transmittance of the input/output coupler. The initial squeezing level in the OPO with 64\,mW pump power can be calculated as $V_\pm=\pm5.82\,$dB and the total losses are $\eta_\text{tot}=0.5$ (see fig.\ref{sqzloss}).\\
Figure \ref{sqzslope} makes use of eq. (\ref{losschannel}) to relate the calculated values for squeezing and anti-squeezing from eq. (\ref{sqzeq1}) and (\ref{sqzeq2}) with the measured values and shows a plot of the detected variance $V^\pm_{\text{det}}$ over the pump power. It is obvious that our novel WPD-lock technique does not degrade the (anti-)squeezing value.
\subsection{Stability}
\begin {figure}[H]
	\centering\includegraphics[width=12cm]{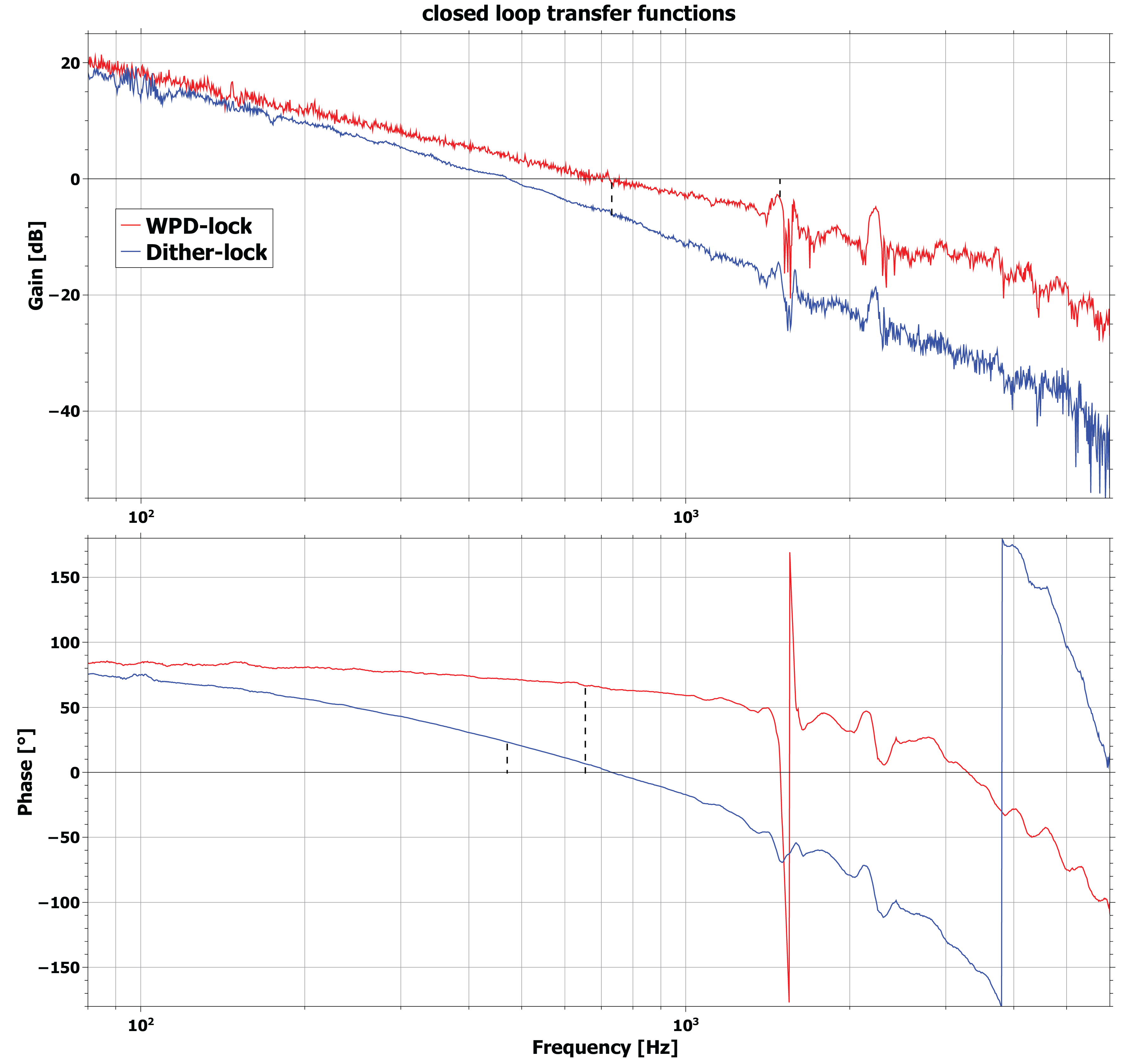}
	\caption{Transfer functions for dither- (blue) and WPD-lock (red). The larger amplitude of the resonance frequencies at 1.7 and 2.2\,kHz is based on the high sensitivity of the photodiodes in the homodyne detector for the WPD-lock.}
	\label{transf}
	\end {figure}

The stability of the system is shown in the transfer functions of the system stabilized via WPD- and dither-lock, respectively (Fig. \ref{transf}). Unity gain frequency is 700\,Hz for WPD- and 450\,Hz for dither-locking of the pump phase. The resonance frequencies that are observable in figure \ref{transf} arise from the piezo-actuated mirror of the OPO-resonator.
	\noindent
The Bode plots show a gain margin of -3\,dB and a phase margin of 70\,$^\circ$ for WPD-lock. The gain and phase margin for dither lock is -7\,dB and 20\,$^\circ$. 
\section{Conclusion}
We have demonstrated a squeezed light source fully stabilized without degradation of the outcoupled squeezed light field. To lock the pump phase to the OPO cavity we used a generally neglected effect called weak pump depletion. We have shown theoretically that there is an influence of the interaction between seed and pump field in the nonlinear medium on every possible outcoupling port of the cavity. However, by using WPD to generate an error signal for pump-phase locking, the detected squeezed states experience no degradation.
\section*{Acknowledgements}
This work was supported financially by the Deutsche Forschungsgemeinschaft through the Centre for Quantum Engineering and Space-Time Research (QUEST) and by the Australian Research Council, grant number DP1094650. T.D. would like to thank Dr. Moritz Mehmet for many helpful discussions and insightful comments.
\end{document}